\documentclass[aps,prd,preprint,superscriptaddress]{revtex4-1}

\usepackage{latexsym}
\usepackage{amsmath,amssymb}
\usepackage{graphicx}
\usepackage{subfigure}
\usepackage{xcolor}
\usepackage{hyperref}     
\hypersetup{colorlinks,%
  citecolor=blue,%
  linkcolor=cyan,%
  pdftex}

\begin{document}


\title{Emergent Friedmann equation from the evolution of cosmic space revisited}

\author{Myungseok Eune} %
\email[]{younms@sogang.ac.kr} %
\affiliation{Research Institute for Basic Science, Sogang University,
  Seoul, 121-742, Republic of Korea} %

\author{Wontae Kim} %
\email[]{wtkim@sogang.ac.kr} %
\affiliation{Research Institute for Basic Science, Sogang University,
  Seoul, 121-742, Republic of Korea} %
\affiliation{Department of Physics, Sogang University, Seoul 121-742,
  Republic of Korea} %
\affiliation{Center for Quantum Spacetime, Sogang University, Seoul
  121-742, Republic of Korea} %

\date{\today}

\begin{abstract}
  Following the recent study on the emergent Friedmann equation from
  the expansion of cosmic space for a flat universe, we apply this
  method to a nonflat universe, and modify the evolution equation to
  lead to the Friedmann equation.  In order to maintain the same form
  with the original evolution equation, we have to define the
  time-dependent Planck length, which shows that the spatial curvature
  of $k=0$ and $k=1$ is preferable to $k=-1$ since the Planck length
  of the nonflat open universe is divergent. Finally, we discuss its
  physical consequences.

\end{abstract}


\maketitle


The thermodynamic quantities of black holes such as its temperature
and entropy are related to geometrical quantities such as surface
gravity and horizon area~\cite{Wald:1984rg}. It has been shown that
the first law of thermodynamics leads to Einstein's field equations
for an accelerating observer~\cite{Jacobson:1995ab}, and then has been
proposed that gravity can be interpreted as an entropic force caused
by changes of entropy associated with the information on a holographic
screen~\cite{Verlinde:2010hp}. By applying the holographic principle
and the equipartition rule of energy, the Einstein field equations
could be derived. These make us know that gravity is an emergent
phenomenon.

Recently, the Friedmann equation governing the
Friedmann-Robertson-Walker (FRW) universe has been remarkably derived
by Padmanabhan~\cite{Padmanabhan:2012ik} from the expansion of cosmic
space due to the difference between the degrees of freedom on a bulk
and in its boundary.  It has been proposed that in an infinitesimal
interval $dt$ of cosmic time, the increase $dV$ of the cosmic volume
in the flat universe is given by
\begin{align}
  \label{dV:dt:Pad}
  \frac{dV}{dt} &= \ell_{\rm P}^2(N_{\rm sur} - N_{\rm bulk}),
\end{align}
where $\ell_{\rm P}$ is the Planck length and $N_{\rm sur}$ and $N_{\rm
  bulk}$ are the degrees of freedom on the surface and in the bulk,
respectively, and the boundary of the bulk is characterized by the
Hubble radius. The relation~\eqref{dV:dt:Pad} yields the standard
dynamical equation in the Friedmann model only for the accelerating
phase of the universe given by $\rho + 3p < 0$, where $\rho$ and $p$
are the energy density and the pressure of a perfect fluid,
respectively. In order to obtain the dynamical equation for any
phases, Eq.~\eqref{dV:dt:Pad} was extended to $dV/dt= \ell_{\rm P}^2
(N_{\rm sur} - \epsilon N_{\rm bulk}) = \ell_{\rm P}^2 (N_{\rm sur} +
N_{\rm m} - N_{\rm de})$, where $N_{\rm m}$ and $N_{\rm de}$ are the
numbers of degrees of freedom of matter with $\rho + 3p >0$ and dark
energy with $\rho + 3p <0$ in the bulk, respectively, and $\epsilon =
+1$ if $\rho + 3p <0$ and $\epsilon =-1$ if $\rho + 3p >0$.
Now, applying the continuity equation to the dynamical equation
derived from Eq.~\eqref{dV:dt:Pad}, the ($n+1$)-dimensional Friedmann
equation can be nicely obtained for the nonflat universe by taking
into account the integration constant which plays a role of the
spatial curvature~\cite{Cai:2012ip}.
There is another generalization of the emergence of cosmic space for a
($n+1$)-dimensional FRW universe corresponding to general relativity,
Gauss-Bonnet gravity, and Lovelock gravity~\cite{Yang:2012wn}. They
proposed that the dynamical equation~\eqref{dV:dt:Pad} was generalized
as \(dV/dt = \ell_{\rm P}^2 f(\Delta N, N_{\rm sur} )\) with \(\Delta
N = N_{\rm sur} - N_{\rm bulk}\).

On the other hand, there have been extensive studies for
thermodynamics in cosmology~\cite{Bak:1999hd, Cai:2005ra,
  Kim:2008zt,Cai:2008gw}. In the nonflat universe, the thermodynamical
quantities are related to the apparent horizon instead of the Hubble
radius; that is, the corresponding temperature is given by the surface
gravity on the apparent horizon and the entropy is proportional to the
area of the apparent horizon. The thermodynamical quantities
associated with the apparent horizon satisfy the first law of
thermodynamics.
In this regard, the apparent horizon is carefully considered to obtain
the Friedmann equation so that $\ell_{\rm P}^2$ in
Eq.~\eqref{dV:dt:Pad} should be replaced by $\ell_{\rm P}^2
\tilde{r}_A/H^{-1}$ in Ref.~\cite{Sheykhi:2013xga}, where
$\tilde{r}_A$ and $H$ are the apparent horizon and the Hubble
parameter, respectively.  However, the bulk is still regarded as a
sphere in the Euclidean space even for the nonflat space, so that the
volume is calculated as the volume of the sphere with the radius of
the apparent horizon.
 
In this brief report, we would like to extend the evolution equation
by taking into account the appropriate invariant volume corresponding
to the nonflat space instead of the volume of a sphere for the flat
space.  We will maintain the form of Padmanabhan's evolution equation
that the expansion of the universe is due to the difference from the
degrees of freedom in the holographic surface between those in the
emerged bulk. Then, the Planck length in the original evolution
equation can be simply redefined by replacing the time-dependent
Planck length.  Finally, we will discuss its physical consequence.


Now, let us consider a spatially homogeneous and isotropic spacetime
given by the line element of
\begin{align}
  ds^2 = h_{ab} dx^a dx^b + \tilde{r}^2 (d\theta^2 + \sin^2 \theta
  d\phi^2), \label{metric}
\end{align}
where $\tilde{r} = a(t) r$, $x^0 = t$, $x^1 = r$, and $h_{ab} =
\mathrm{diag}(-1, a^2/(1-kr^2))$. Here, $k$ denotes the curvature of
space with $k=-1$, $0$, and $1$, corresponding to open, flat, and
closed universes, respectively. The apparent horizon can be calculated
from the relation $\left. h^{ab} \partial_a \tilde{r}\partial_b
  \tilde{r} \right|_{\tilde{r} = \tilde{r}_A} = 0$ and is obtained as
\begin{align}
  \tilde{r}_A = \frac{1}{\sqrt{H^2 + k/a^2}}, \label{r:A}
\end{align}
where the Hubble parameter is given by $H = \dot{a}/a$ and the overdot
denotes the derivative with respect to comoving time $t$. Note that
the apparent horizon becomes $\tilde{r}_A = H^{-1}$ for flat space
with $k=0$.  We assume that the number of degrees of freedom on the
surface at the apparent horizon is proportional to its area $4\pi
\tilde{r}_A^2$ and is given
by~\cite{Padmanabhan:2012ik,Cai:2012ip,Sheykhi:2013xga}
\begin{align}
  \label{N:sur}
  N_{\rm sur} = 4S = \frac{4\pi \tilde{r}_A^2}{\ell_{\rm P}^2},
\end{align}
where $S$ is the entropy. For simplicity, we set $k_\mathrm{B} = c =
\hbar = 1$. 
The Hawking temperature associated with the apparent horizon is given
by~\cite{Cai:2005ra,Sheykhi:2013xga}
\begin{align}
  \label{HawkingTemperature}
  T_{\rm H} = \frac{1}{2\pi \tilde{r}_A}.
\end{align}
At a given time, an invariant volume of space surrounded by the
apparent horizon can be written as
\begin{align}
  \label{V:k}
  V_k = 4\pi a^3 \int_0^{\tilde{r}_A/a} dr \frac{r^2}{\sqrt{1-kr^2}},
\end{align}
which reduces to $V_0 = 4\pi \tilde{r}_A^3/3$ for $k=0$ and $V_k =
2\pi a^2 \left[ \sqrt{k}\, a \sin^{-1} \left(\sqrt{k}\, \tilde{r}_A/a
  \right) - k \tilde{r}_A^2 H \right]$ for $k=\pm 1$.  Next, the Komar
energy within the bulk~\cite{Padmanabhan:2012ik, Cai:2012ip,
  Sheykhi:2013xga} is modified by
\begin{align}
  E_k = (\rho + 3p) V_k, \label{E:k}
\end{align}
where it depends on the spatial curvature $k$. Using the equipartition
rule of energy, the degrees of freedom in the bulk can be written as
\begin{align}
  N_{\rm bulk} = \frac{2|E_k|}{T_{\rm H}}. \label{N:bulk:def}
\end{align}

We assume that in an infinitesimal interval $dt$, an increase $dV_k$
of the invariant volume even in the nonflat FRW universe including the
flat universe is still proportional to the difference between $N_{\rm
  sur}$ and $N_{\rm bulk}$ as~\cite{Padmanabhan:2012ik}
\begin{align}
  \frac{dV_k}{dt} = \ell_{\rm P}^2 f_k(t) (N_{\rm sur} - \epsilon N_{\rm
  bulk}), \label{dV:dt}
\end{align}
where $\epsilon N_{\rm bulk}$ is given by $ \epsilon N_{\rm bulk}=
N_{\rm de} - N_{\rm m} = - (2V_k/T_{\rm H}) (\rho + 3p)$ and $f_k (t)$
is a proportional function which can be chosen appropriately to give
the Friedmann equation such as
\begin{align}
  f_k(t) = \frac{\bar{V}_k \left[\dot{\tilde{r}}_A H^{-1} /\tilde{r}_A
      + (\tilde{r}_A/H^{-1}) (H^{-1}/\tilde{r}_A - V_k/\bar{V}_k)
    \right]}{V_k \left(\dot{\tilde{r}}_A H^{-1} /\tilde{r}_A +
      \bar{V}_k/V_k - 1\right)}, \label{ratio:f}
\end{align}
with $\bar{V}_k \equiv 4\pi \tilde{r}_A^3/3$.  We see that $f_0(t) =1$
for the flat universe with $k=0$, and so it is compatible with the
previous results~\cite{Cai:2005ra,Sheykhi:2013xga}.

To show the better motivation for Eq.~\eqref{ratio:f} and discuss the
difference from the previous results, we have to mention the work done
in Ref.~\cite{Sheykhi:2013xga}. The extension to a nonflat space in
this cosmic model has been performed by carefully treating the general
expression of the apparent horizon depending on the spatial
curvature. For this reason, the original equation~\eqref{dV:dt:Pad}
should have been modified so that the author was able to derive the
FRW universe with any spatial curvatures. The key ingredient is to
assume the nontrivial proportional function instead of the
proportional constant in order to get the correct FRW equation even
for the nonflat universe. However, the volume of the universe was
still defined in the flat space rather than the general expression of
$V_k$ defined for the nonflat universe. All physical quantities
reflected the nontrivial dependence of $k$ except for the volume in
Ref.~\cite{Sheykhi:2013xga}. It shows that the time evolution of the
flat universe generates the nonflat FRW equation. To avoid this
interpretation, we have started with the nonflat universe from the
beginning. That is the reason why we have considered the volume
increase of the nonflat universe such as Eq.~\eqref{dV:dt} along with
Eq.~\eqref{ratio:f}.

By taking the time derivative of the invariant volume~\eqref{V:k}, we
obtain
\begin{align}
  \frac{dV_k}{dt} &= 4\pi \tilde{r}_A (\dot{\tilde{r}}_A H^{-1} -
  \tilde{r}_A) + 3H V_k \notag \\
  &= 4\pi \tilde{r}_A^2 \left[\frac{\dot{\tilde{r}}_A
      H^{-1}}{\tilde{r}_A} + \frac{\tilde{r}_A}{H^{-1}}
    \left(\frac{H^{-1}}{\tilde{r}_A} - \frac{V_k}{\bar{V}_k} \right)
  \right]. \label{dV:dt:cal}
\end{align}
The relation between the degrees of freedom is calculated as
\begin{align}
  N_{\rm sur} - \epsilon N_{\rm bulk} &= \frac{4\pi
    \tilde{r}_A^2}{\ell_{\rm P}^2} + 4 \pi \tilde{r}_A (\rho + 3p)
  V_k. \label{Delta:N}
\end{align}
Eliminating $p$ in Eq.~\eqref{Delta:N} by the use of the continuity
equation $\dot\rho + 3H (\rho + p) = 0$, we can obtain
\begin{align}
  N_{\rm sur} - \epsilon N_{\rm bulk} &= \frac{4\pi
    \tilde{r}_A^2}{\ell_{\rm P}^2} \left[1 - \frac{\ell_{\rm P}^2
      V_k}{H \tilde{r}_A} (\dot\rho + 2\rho H)
  \right] \notag \\
  &= \frac{4\pi \tilde{r}_A^2}{\ell_{\rm P}^2} \left[1 -
    \frac{\ell_{\rm P}^2 V_k}{H\tilde{r}_A a^2} \frac{d}{dt}
    (\rho a^2) \right]. \label{Delta:N:cal}
\end{align}
Substituting Eqs.~\eqref{ratio:f}, \eqref{dV:dt:cal},
and~\eqref{Delta:N:cal} into Eq.~\eqref{dV:dt}, we get
\begin{align}
  \frac{d}{dt} \left(\frac{a^2}{\tilde{r}_A^2} \right) = \frac{d}{dt}
  \left[a^2 \left(H^2 + \frac{k}{a^2} \right) \right] = \frac{8\pi
    \ell_{\rm P}^2}{3} \frac{d}{dt}(\rho a^2). \label{dV:dN}
\end{align}
Integrating Eq.~\eqref{dV:dN}, one can obtain
\begin{align}
  H^2 + \frac{k}{a^2} = \frac{8\pi \ell_{\rm P}^2}{3}
  \rho \label{Friedmann}.
\end{align}
Note that one can regard the integration constant as the curvature of
space $\tilde{r}_A = 1/H$~\cite{Cai:2012ip}, while one can set the
integration constant to zero for $\tilde{r}_A = 1/\sqrt{H^2 +
  k/a^2}$~\cite{Sheykhi:2013xga}.  In the present case, we can take
the vanishing integration constant for simplicity.

It is interesting to note that the original form of the evolution
equation~\eqref{dV:dt:Pad} can be maintained if we replace the
proportional function along with the Planck length by the effective
Planck length as $\ell_{\rm P}^2 f_k(t) = (\ell_{\rm P}^2)^{\rm
  eff}_{k}$ in the modified equation \eqref{dV:dt}. What it means is
that the Newton constant may be running depending on the curvature of
space as long as one can take the square of the Planck length as a
gravitational coupling$ $.  In particular, the Newton constant is
merely constant for the flat universe of $k=0$.  In order to examine
the late time behavior of the effective Planck length, we now
consider three cases: the radiation-dominant, the matter-dominant, and
the $\Lambda$-dominant universes, where $\Lambda$ denotes the positive
cosmological constant.
First, the radiation-dominant universe is described by the scale
factor $a=a_0 t^{1/2}$, where $a_0$ is a positive constant.  The
square of the effective Planck length goes to $ (\ell_{\rm P}^2)^{\rm
  eff}_{k=-1} \approx \ell_{\rm P}^2 (1/3) (1/2 - t^{1/2}/a_0)^{-1/2}
$ with $t<a_0^2/4$.  At $t = a_0^2/4$, it is divergent for the open
universe of $k=-1$ while it is asymptotically constant since
$(\ell_{\rm P}^2)^{\rm eff}_{k=1} =\ell_{\rm P}^2[2 - (3\pi a_0/8)
t^{-1/2} + O\left(t^{-1} \right)]$ for the closed universe of $k=1$.
Second, in the matter-dominant universe described by $a=a_0 t^{2/3}$,
the square of effective Planck length behaves as $(\ell_{\rm P}^2)^{\rm
  eff}_{k=-1} = \ell_{\rm P}^2 [(4/9)(4/9-t^{2/3}/a_0^2)^{-1/2} +
O((4a_0^2/9-t^{2/3})^{1/2}) ]$ with $t<(2a_0/3)^3$, and it is
divergent at $t=(2a_0/3)^3$. However it becomes a constant for the
closed universe of $k=1$ since $(\ell_{\rm P}^2)^{\rm eff}_{k=1} =
\ell_{\rm P}^2 [2 - (\pi a_0/2) t^{-1/3} + O(t^{-2/3})]$ at the late
time.  Thus for the radiation- and matter-dominant universes, the
effective Planck length is divergent for $k=-1$ while it approaches
$(\ell_{\rm P}^2)^{\rm eff}_{k=1} = 2 \ell_{\rm P}^2$ for $k=1$.
Finally, the $\Lambda$-dominant nonflat universe is governed by the
scale factor $a=a_0 e^{\alpha t}$ with $\alpha \equiv \sqrt{
  \Lambda/3}$.  The square of the effective Planck length behaves as
$(\ell_{\rm P}^2)^{\rm eff}_{k=\pm 1} = \ell_{\rm P}^2 [12/7+(61
/(343\alpha^2 a_0^2))e^{-2\alpha t} + O(e^{-4\alpha t}) ]$, and both
of them approach $(\ell_{\rm P}^2)^{\rm eff}_{k} = (12/7) \ell_{\rm
  P}^2 $ at the asymptotic infinity. Therefore, for the radiation- and
matter- dominant cases, the Newton constant can be divergent for the
nonflat open universe while it becomes the constant for the nonflat
closed universe.  In the vacuum-energy-dominant case, the Newton
constant can be finite at $k=\pm 1$ and $k=0$.  It shows that if the
nonflat open universe evolves eternally without encountering the
vacuum-energy era, then it undergoes the divergent gravitational
interaction.

As seen from the above calculations, the gravitational coupling
diverges for the nonflat spaces, so that it is natural to ask why the
present calculation is not compatible with the observations. Before
the big bang, all forces were expected to be unified, and then the
gravitational force was decoupled from three forces at $10^{19}$\,GeV
where the temperature was about $10^{32}$\,K. It has been claimed that
the radiation-dominant era, matter-dominant era, and the recent
vacuum-energy-dominant era, all can be described in terms of
thermodynamics by assuming a thermal or quasithermal state of our
universe. However, we found some deviations from the standard results,
and the worst case is that the severely divergent gravitational
coupling at $k=-1$ appears for these eras. The essential drawback of
our formulation is due to the time-dependent
temperature~\eqref{HawkingTemperature} and equipartition
law~\eqref{N:bulk:def}. Note that these have something to do with the
assumption of thermal equilibrium. Therefore, we think that this
incompatibility with the observations should be related to the
validity of the temperature~\eqref{HawkingTemperature} and
equipartition law~\eqref{N:bulk:def} when we consider the nonflat
universe. Of course, the best fit appears at the $\Lambda$-dominant
era of the flat universe since the Hawking
temperature~\eqref{HawkingTemperature} is no more time dependent,
which is definitely constant, so that we can use the equipartition law
and the related thermal quantities in equilibrium, which is compatible
with the qualitative behavior of the accelerated expansion of the
universe with $k=0$. As a result, we have tried to extend the original
idea that the cosmic expansion appears from the difference of the
degrees of freedom on the bulk and that of the boundary to the nonflat
space; however, it is not easy to obtain meaningful interpretations
except for the flat universe since the time-dependent
temperature~\eqref{HawkingTemperature} and the equipartition
law~\eqref{N:bulk:def} can be inappropriate to the nonflat universe.

In conclusion, following the proposal that the space evolution is due
to the different degrees of freedom on the holographic surface and its
bulk, we have extended the evolution equation to give the Friedmann
equation even in the nonflat universe corresponding to $k=\pm 1$ by
taking into account the invariant volume surrounded by the apparent
horizon.  Moreover, the limit to the flat universe of $k=0$ can be
easily recovered.


\begin{acknowledgments}
  We would like to thank Yongwan Gim, Edwin J. Son, and Sang-Heon Yi
  for exciting discussions.  This work was supported by the Sogang
  University Research Grant of (2013) 201310022.
\end{acknowledgments}


\bibliographystyle{apsrev4-1} 
\bibliography{references}


%
%

\end{document}